

\overfullrule = 0pt\def\frac#1#2{{#1 \over #2}}
\def\inv#1{{1 \over #1}}
\def\half{{1 \over 2}}
\def\dd#1#2{{\partial #1 \over \partial #2}}
\def\Gl{\lambda}\def\Gth{\theta}

\magnification=1200
\rightline{hepth@xxx/9201055}
\centerline{{\bf Addendum to}}
\medskip\centerline{{\bf Combinatorics of the Modular Group II }}
\smallskip\centerline{{\bf The Kontsevich Integrals }}
\smallskip\centerline{hepth@xxx/9201001}
\vskip6mm\centerline{{ C. Itzykson and J.-B. Zuber }}
\vskip10mm
At the end of the paper, after equation (7.18),
replace ``From this system \dots  in sec.
5 and 6 \dots'' by\par
\vskip10mm
The resulting equation may be expanded for large 
$\lambda_j$
and the coefficients of successive inverse powers,
expressed as differential operators in the variables
$\Gth_k= \inv{k} \sum_i \Gl_i^{-k}$,  set equal to zero.
For our purpose, it is sufficient to consider the two leading terms of
order $\Gl^{-2}$  and $\Gl^{-3}$. Surprisingly, the coefficient of the latter
depends on the variables $\Gth_{3n}$, of which $Z$ itself
is independent, {\it v.i.z.}
$$ {\rm l.h.s.\ of\ }(7.17)=\inv{\Gl^2} L_{-1} Z +
\inv{\Gl^3}\left(-W_{-2}-2 \sum_{n\ge 1} n\Gth_{3n} L_{n-2}\right)Z
+{\rm O}\left(\inv{\Gl^4}\right)\ . \eqno(7.19)$$
%
%
%
As the notation suggests, the coefficients are
generators of the $W_3$ algebra. Their explicit form reads
$$\eqalignno{\qquad L_{n} & =\inv{3}\sum_{}(k-3n) \Gth_{k-3n}\dd{}{\Gth_k}
+\inv{6}\sum_{k+l=3n}\dd{}{\Gth_k} \dd{}{\Gth_l}\qquad \cr
& {}-\half  \dd{}{\Gth_{3n+4}} +\frac{2}{3}\Gth_1\Gth_2 \delta_{n,-1}
+\inv{9}\delta_{n,0}
\qquad \qquad n\ge -1&(7.20)\cr }$$
and
$$\eqalignno{ W_{-2}& = \inv{9}\sum
\left( (k+l+6) \Gth_{k+l+6}
\dd{}{\Gth_k}\dd{}{\Gth_l}+kl \Gth_k\Gth_l\dd{}{\Gth_{k+l-6}}\right) \cr
& {}
-\inv{3}\sum
(k+2)\Gth_{k+2}\dd{}{\Gth_k} +\inv{4}\dd{}{\Gth_2}
-\inv{6} \Gth_1^2+\frac{4}{9}\Gth_1^2\Gth_4 +\frac{8}{27}\Gth_2^3\
&(7.21) \cr}$$
where it is understood that terms where the subscript of a $\Gth$ is
negative or a multiple of 3 are absent. By the commutation relations of
the $W_3$ algebra, $[L_n, W_{-2}]=2(n+1)W_{n-2}$,  one constructs
$$\eqalignno{
W_{n}& = \inv{9}\sum
\left( (k+l-3n) \Gth_{k+l-3n}
\dd{}{\Gth_k}\dd{}{\Gth_l}+kl \Gth_k\Gth_l\dd{}{\Gth_{k+l+3n}}\right) \cr
& {}
-\inv{3}\sum
(k-3n-4)\Gth_{k-3n-4}\dd{}{\Gth_k}
+\inv{27}\sum_{j+k+l=3n}\dd{}{\Gth_j}\dd{}{\Gth_k}\dd{}{\Gth_l}\cr
& {} -\inv{6}\sum_{k+l=3n+4}\dd{}{\Gth_k}\dd{}{\Gth_l}+\inv{4}
\dd{}{\Gth_{3n+8}} & (7.22) \cr
& {}
+\left(\frac{4}{9}\Gth_1^2\Gth_4 +\frac{8}{27}\Gth_2^3\ -\inv{6} \Gth_1^2
\right) \delta_{n,-2}+\inv{27}\Gth_1^3\delta_{n,-1} \ ,
\qquad\qquad n\ge -2\ . \cr
}$$
Up to a shift of the variable $\Gth_4$ by $-3/8$,
these expressions coincide with
those given by Goeree 
[16]. A similar calculation can presumably be carried
out in the general case of $W_p$, where one would study an expansion of
the generalized matrix Airy equation, keeping terms of order
$\Gl^{-2},\Gl^{-3},\cdots,\Gl^{-p}$.

Returning to eq. (7.19), the equations $L_n Z=W_m Z=0$, ($n+1, m+2\ge 0$,
or rather the generating set $\{-1\le n\le 2, m=-2\}$), determine $Z=e^F$.
Alternatively one may use the leading term $L_{-1}Z=0$ (the so-called
string equation) and the KdV${}_3$ hierarchy
$$\eqalignno{ \dd{Q}{\Gth_s}& = [Q^{{s\over 3}}_+,Q] \cr
Q & = \left( \dd{}{\Gth_1}\right)^3 +\frac{3}{2} \left\{u_1,\dd{}{\Gth_1}
\right\} +3u_2 \cr
u_1 & = \dd{{}^2F}{\Gth_1^2}\qquad\qquad u_2=\half
{\partial^2 F\over \partial \Gth_1\partial\Gth_2} \ . & (7.23) \cr
}$$
Finally it is worth pointing out the relation between the $\Gth$-variables
of the matrix model and those used by Witten [8] 
in the topological
context. For the case $p=3$ ($W_3$ algebra) he introduces variables
$t_{n,m}$ where $n\ge 0$ and $m=0,1$ with
$$\eqalignno{t_{n,m} & ={(3n+m+1)!!!\over \rho^{3n+m+1}}\left(
{3^{\half}\over i}\right)^{n+m} \Gth_{3n+m+1}\cr
\rho^4 & = {3^{{3\over 2}}\over 2i}& (7.24) \cr
(3n+m+1) !!!&\equiv (3n+m+1)(3(n-1)+m+1)\cdots(3+m+1)(m+1) \ . 
\cr}$$
The factors $i$ disappear from the final expressions. Indeed there is a
topological constraint expressing that in $F$ the only non-vanishing
terms $\prod_k t_{n_k,m_k}$ satisfy
$$ 8(g-1)=\sum_k[(3n_k+m_k+1)-4] \eqno (7.25) $$
with $g$ the corresponding genus. Hence such a term translates into
$$ \prod_k t_{n_k,m_k} = (-12)^{g-1} {(-2)^{\Sigma_k\, 1}\over
(-3)^{\Sigma_k\, n_k}}
\prod_k (3n_k+m_k+1)!!!\, \Gth_{3n_k+m_k+1}\ .\eqno (7.26)$$
For instance the leading part of $F$ restricted to genus $0$ and to the
subspace where only $\Gth_1$, $\Gth_2$ are non-vanishing (the ``small phase
space'') reads
$$ f={2\over 3}\Gth_1^2 \Gth_2 -{8\over 27} \Gth_2^4 ={t_{0,1}^2 t_{0,2}
\over 2} +{t_{0,2}^4\over 3.4!}\ . \eqno(7.27)$$
Further generalization of Kontsevich integrals to other hierarchies
of integrable flows ($D$ or $E$ series) remains a challenging problem.
\vskip15mm
We have received recently two papers that overlap with this work
and complement our presentation
[17][18].

\vskip2cm
\centerline{{\bf Additional references}}
\medskip\noindent[16]\quad J. Goeree,
{\it $W$ constraints in 2-D quantum gravity}, Nucl. Phys. {\bf B358}
 (1991) 737-757.\par\noindent
[17]\quad R. Dijkgraaf, {\it Intersection theory, integrable hierarchies
and topological field theory}, Princeton preprint IASSNS-HEP-91/91,
hepth@xxx/9201003.
\par\noindent[18]\quad S. Kharchev, A. Marshakov, A. Mironov, A. Morozov and
A. Zabrodin, {\it Towards unified theory of 2d gravity},
ITEP-Lebedev Institute preprint FIAN/TD-10/91, hepth@xxx/9201013.

\bye